\documentclass[11pt]{article}
\makeatletter
\def\section{\@startsection{section}{1}{\z@}{-3.5ex plus -1ex minus -2.ex}
{2.3ex plus .2ex}{\Large\bf}}

\def\subsection{\@startsection{subsection}{2}{\z@}{-3.25ex plus
 -1ex minus -2.ex}
{1.5ex plus .2ex}{\bf}}

\def\vsn{\vskip 1pc \noindent}
\def\f{\newline}
\def\e{\varepsilon}

\def\comp{{\rm comp}}

\def\rr{{\bf R}}
\def\nn{{\bf N}}

\def\P{{\rm{\bf P}}}

\newcommand{\be} {\begin{equation}}
\newcommand{\ee} {\end{equation}}
\newcommand{\bd} {\begin{displaymath}}
\newcommand{\ed} {\end{displaymath}}

\newcommand{\bq}{\begin{eqnarray}}
\newcommand{\eq}{\end{eqnarray}}
\newcommand{\bqn}{\begin{eqnarray*}}
\newcommand{\eqn}{\end{eqnarray*}}
\newcommand{\ba}[1]{\begin{array}{#1}}
\newcommand{\eqa}{\end{array}}
\def\qed{
   \\[-4ex]
  \hbox to \hsize{\hfill \vrule height 1.6ex width 1.5ex
  depth -.1ex}}

\begin{document}

\bibliographystyle{alpha}

\begin{center} {\Large {\bf
Almost Optimal Solution of Initial-Value Problems by Randomized and 
Quantum Algorithms 
 }\footnotemark[1] } 
\end{center}
\footnotetext[1]{ ~\noindent This research was partly supported by  
  AGH grant No. 10.420.03 \vsn}               
  
\medskip
\begin{center}
{\large {\bf Boles\l aw Kacewicz \footnotemark[2] }}
\end{center}
\footnotetext[2]{ 
\begin{minipage}[t]{16cm} 
 \noindent
{\it Department of Applied Mathematics, AGH University of Science 
and Technology,\\
\noindent  Al. Mickiewicza 30, paw. A3/A4, III p., 
pok. 301,\\
 30-059 Cracow, Poland 
\newline
 kacewicz@uci.agh.edu.pl, tel. +48(12)617 3996, fax +48(12)617 3165 }  
\end{minipage} }

\thispagestyle{empty}
$~$
\vsn
\vsn
\vsn
\vsn
 
We establish essentially optimal bounds on the complexity of
initial-value problems in the randomized and quantum settings. 
For this purpose we define a sequence of new algorithms whose
error/cost properties improve from step to step. 
These algorithms yield new upper complexity bounds,
which differ from known lower bounds by only an arbitrarily small
positive parameter in the exponent, and a logarithmic factor. In both the
randomized and quantum settings, initial-value problems turn out 
to be essentially as difficult as scalar integration. 
\newpage
\noindent
{\Large \section{ Introduction }}
\noindent
 The complexity of initial-value problems has been well understood for many years
 in the deterministic settings, see \cite{przegl} for a survey of the worst-case and asymptotic
results.  
 Much less is known about the complexity of these problems if randomized methods 
or quantum computation are allowed. Other  problems related to initial-value problems, 
such as summation or integration, have already been well studied in both randomized
and quantum settings. A survey of complexity results for summation and integration 
can be found in \cite{HN}.
\f
Randomized and quantum complexity  of initial-value problems was first studied in \cite{rand5}. 
This paper showed that a nontrivial 
speed-up can be achieved by switching from the worst-case deterministic setting
to the randomized or quantum computation. Improved  complexity bounds, 
the best known up to now, were next shown in  \cite{new}. 
In the H\"older class of right-hand side functions with $r$ continuous 
bounded partial derivatives, with the $r$th derivative being
a H\"older function with exponent $\rho$, 
the $\e$-complexity was shown to be 
$O\left( (1/\e)^{1/(r+\rho+1/3)} \right)$ in the randomized setting, 
and $O\left( (1/\e)^{1/(r+\rho+1/2)} \right)$ on a quantum computer
(logarithmic factors are here neglected). 
\f
Unfortunately, these upper bounds
do not match  known lower bounds on the $\e$-complexity. It can be shown
 \cite{rand5} that the $\e$-complexity is 
$\Omega\left( (1/\e)^{1/(r+\rho+1/2)} \right)$ in the randomized case, 
and $\Omega\left( (1/\e)^{1/(r+\rho+1)} \right)$ on a quantum computer. 
Hence, aside from logarithmic factors, there is a significant difference between the bounds
in the exponent. 
The gap was essentially filled for 
a restricted class of scalar autonomous problems in \cite{new}. 
For systems of equations, the discrepancy between upper and lower 
bounds remained an open problem. These results are briefly
recalled in Section 2.
\f
In this paper, we show almost optimal upper complexity bounds 
in the randomized and quantum settings for 
a general initial-value problem. 
The main results are contained in Theorems~1 and~2. We recursively define a sequence
$A_1, A_2,\ldots, A_k$ of randomized or quantum algorithms (which are new for $k\geq 3$)
for systems of initial-value problems. The performance of $A_k$ improves as $k$ grows, 
which is a crucial component of the complexity analysis. 
The error and cost bounds for~$A_k$ are shown in Theorem~1.
\f
Properties of $A_k$, with $k$ properly chosen, 
lead to new upper complexity bounds given in Theorem~2. These differ from the 
lower bounds by 
only an arbitrarily small positive parameter $\gamma$ in the exponent, 
and a logarithmic factor. The $\e$-complexity turns out to be
$O\left( (1/\e)^{1/(r+\rho+1/2-\gamma)} \right)$ in the randomized setting, 
and $O\left( (1/\e)^{1/(r+\rho+1-\gamma)} \right)$ on a quantum computer,
up to a logarithmic factor.
Hence, in both the randomized and quantum settings, the complexity of 
initial-value problems is essentially established. 
Roughly speaking, the problem is as difficult as scalar integration \cite{HN}. 
\f
{\Large \section{ Problem definition and known complexity bounds }}
\noindent
In this section we define the problem to be solved, and briefly introduce 
basic notions  on randomized and quantum settings used in the rest 
of the paper.
We also recall, for later comparison, known results 
on the complexity of initial-value problems in both settings.
\vsn
We consider the solution of a system of ordinary 
differential equations with initial conditions
\be
z'(t)=f(z(t)), \;\;\; t\in [a,b],\;\;\;\;\; z(a)=\eta,
\label{1}
\ee
where $f:\rr^d \to \rr^d$, $z:[a,b]\to \rr^d$ and $\eta\in \rr^d\, $ 
($f(\eta)\ne 0$). 
Nonautonomous systems 
$z'(t)=f(t,z(t))$ with  $f: \rr^{d+1} \to \rr^d$
can also be written in the form (\ref{1}), by adding one 
scalar equation
$$ \left[ \begin{array}{l} u'(t) \\ z'(t)\end{array} \right]
= \left[ \begin{array}{c} 1 \\ f(u(t), z(t)) \end{array} \right],
$$
with an additional initial condition $u(a)=a$. 
\vsn
We shall study the complexity of problem (\ref{1}) in the randomized 
and quantum settings. 
We assume that the right-hand side function $f=[f^1,\ldots,f^d]^T$ 
belongs to the H\"older class $F^{r,\rho}$ given as follows.
For an integer $r\geq 0$, $\rho\in (0,1]$, and 
positive numbers $D_0,D_1,\ldots, D_r$, $H$, we~define
$$
F^{r,\rho} =\{\, f:\rr^d\to \rr^d \mid \; f\in C^r(\rr^d), \;\;\;
|\partial ^i f^j(y)| \leq D_i, \; |i|\leq r,
$$
\be
 |\partial^r f^j(y)-\partial^r f^j(z)|\leq 
H\, \|y-z\|^{\rho}, \; y,z \in \rr^d,\;\; j=1,2,\ldots, d \, \},
\label{2}
\ee
where $\partial^i f^j$ represents any partial derivative of order $|i|$ 
of the $j$th component of $f$. Here, and in the rest of this paper,
$\|\cdot\|$ denotes the maximum norm in $\rr^d$. 
\f
Let $q=r+\rho$ denote the regularity parameter for the class $F^{r,\rho}$.
We assume that $q\geq 1$, which assures that $f$ is always a 
Lipschitz function. 
\vsn
We now briefly recall  
basic notions related to the randomized and quantum settings.
For a~complete discussion, in particular of the
quantum setting, the reader is referred to~\cite{Heinrich}.
\f
We wish to compute a bounded function $l$ on $[a,b]$ approximating
the solution $z$. 
Information available about the right-hand side $f$ in the construction of $l$ 
 is provided by a subroutine that computes values of a component of $f$ 
or one of its partial derivatives. In the randomized setting, 
the points at which the values are computed can be selected randomly. 
On a quantum computer, by subroutine calls we mean 
either applications of a quantum query operator for (a component of) $f$, or
evaluations of  components of $f$ or its partial derivatives on a classical
computer. A~detailed discussion of a quantum query operation can be found
in \cite{Heinrich}. A mapping $A$ that computes $l$ based on the available 
information is called an {\it algorithm}. In both randomized and quantum settings,
the approximating function produced by an algorithm is random.
\f
We now define the error of an algorithm in the two settings.
Let  ($\Omega$, $\Sigma$, $\P$) be a probability space. 
Let $\{x_i\}$ define the uniform partition of $[a,b]$, so that
$x_i=a+ih$, $i=0,1,\ldots,n$ with $h=(b-a)/n$. The approximation $l=l(t)$ 
to $z=z(t)$ in $[a,b]$ is
constructed in the algorithm $A$ based on random approximations 
$y_i=y_i^{\omega}$ to $z(x_i)$, $i=0,1,\ldots,n$, $\omega\in \Omega$. 
(We assume that the mappings $\omega \mapsto y_i^\omega$
are random variables for each $f\in F^{r,\rho}$.) More specifically, $l=l^\omega$
is given by 
\be
l^\omega(t)=\psi(y_0^\omega, y_1^\omega,\ldots, 
y_n^\omega)(t)\, ,
\label{3.2}
\ee
for $t\in [a,b]$, where $\psi$ is a certain mapping defining
an approximation over $[a,b]$ from discrete approximations $y_i^\omega$.
\f
The error of $A$ at $f$ is defined by 
\be
e^\omega(A,f)=\sup_{t\in [a,b]}\|z(t)-l^\omega(t)\|.
\label{3.3}
\ee
We assume that the mapping $\omega \mapsto e^\omega(A,f)$
is a random variable for each $f\in F^{r,\rho}$.
\f
In the {\it randomized setting}, the {\it error} of $A$ in the class $F^{r,\rho}$ is
given by the maximal dispersion of $e^\omega(A,f)$, namely, 
\be
e^{{\rm rand}}(A, F^{r,\rho}) = \sup\limits_{f\in F^{r,\rho}}
({\rm {\bf E}} e^\omega(A,f)^2)^{1/2} \, ,
\label{3a}
\ee
where {\bf E} is the expectation.  
(The maximal expected value of $e^\omega(A,f)$ can be considered as well. 
This only changes constants in the results.)
\f
In the {\it quantum setting},
the {\it error} of $A$ in the class $F^{r,\rho}$ is given probabilistically 
by
\be
e^{{\rm quant}}(A, F^{r,\rho}) =
e^{{\rm quant}}(A, F^{r,\rho},\delta) = \sup\limits_{f\in F^{r,\rho}}
\inf\; \{\; \alpha|\;\; \P\{\, e^{\omega}(A,f)>\alpha \,\}\; 
\leq \delta\; \},
\label{3b}
\ee
for a given number $\delta$, $0<\delta<\frac{1}{2}$.
\vsn
The {\it cost}  of an algorithm $A$, ${\rm cost}(A)$, in the randomized 
or quantum settings is measured by the maximal number  (with respect
to $f\in F^{r,\rho}$)
of subroutine calls (with a proper meaning of a subroutine call
in the respective setting) needed to compute an approximation. 
For a given $\e>0$,  the $\e$-{\it complexity} of the problem in the respective 
setting, $\comp^{{\rm rand(quant)}} (F^{r,\rho}, \e)$, is defined to be 
the minimal cost of an algorithm $A$ taken among all $A$
such that  $e^{{\rm rand(quant)}}(A, F^{r,\rho}) \leq \e$. 
In the quantum setting, to display the dependence on $\delta$, we shall
denote the complexity by $\comp^{{\rm quant}} (F^{r,\rho}, \e, \delta)$. 
\vsn
We now briefly recall known results on the subject. 
For scalar autonomous problems  (\ref{1}) (with $d=1$), matching complexity bounds,
 up to logarithmic factors, were established in \cite{new}.
It was shown, in a slightly restricted class $\hat{F}^{r,\rho}$,
and for a problem restricted to computing $z(b)$ only, that
\be
\comp^{{\rm rand}} (\hat{F}^{r,\rho}, \e) =
O \left( \left( \frac{1}{\e}\right)^{1/(q+1/2)} \,
 \left(\log \frac{1}{\e} \right)^2   \right)
\label{rec1}
\ee
and 
\be
\comp^{{\rm rand}} (\hat{F}^{r,\rho}, \e) =
\Omega \left(\left(\frac{1}{\e} \right)^{1/(q +1/2)} \right) \, .
\label{rec2}
\ee
\f
In the quantum setting
\be
\comp^{{\rm quant}} (\hat{F}^{r,\rho}, \e, \delta) =
O \left(\left( \frac{1}{\e}\right)^{1/(q+1)} 
\left( \log\log \frac{1}{\e} + \log \frac{1}{\delta} \right)
\log \frac{1}{\e}\right) 
\label{rec3}
\ee
and, for $0 < \delta\leq \frac{1}{4}$, 
\be
\comp^{{\rm quant}} (\hat{F}^{r,\rho}, \e, \delta) =
\Omega \left( \left(\frac{1}{\e}\right)^{1/(q +1)} \right) \, .
\label{rec4}
\ee 
The proof of these results  was based on switching 
to an equivalent scalar nonlinear equation. 
Such a method cannot be used for systems of equations.
The best known complexity bounds for a general problem are given in \cite{new}.
It has been shown that 
\be
\comp^{{\rm rand}} (F^{r,\rho}, \e) =
O\left( \left(\frac{1}{\e}\right)^{ 1/(q +1/3) } \log \frac{1}{\e} \right)
\label{rec5}
\ee
and 
\be
\comp^{{\rm quant}} (F^{r,\rho}, \e, \delta) =
O\left( \left( \frac{1}{\e}\right) ^{1/(q+1/2)} 
\left( \log \frac{1}{\e} + \log \frac{1}{\delta}\right)\right)  \, . 
\label{rec6}
\ee 
Similarly to the case of scalar equations, lower bounds are given
by (\ref{rec2}) and (\ref{rec4}).
Hence, there is a significant gap in known estimates of the complexity
for systems of equations.
\f
In the next part of this paper we show that the existing gap can essentially
be filled up. 
\vsn
{\Large \section{ Main results  }}
\noindent
In this section we present main results of this paper.
The first result, Theorem 1, assures the existence of a class of algorithms 
for solving problem (\ref{1})
in the randomized (RAND) and quantum (QUANT) settings, 
which possess certain error and cost 
properties. These properties will allow us to derive the desired 
complexity bounds. The algorithms are defined constructively  
in the next section. 
The next result, Theorem 2, is an immediate consequence of Theorem 1. 
It contains new, and almost optimal, complexity bounds for problem (\ref{1}) 
in the randomized and quantum settings.
\f
In what follows, $L$ denotes the Lipschitz constant for $f$. 
\vsn
{\bf Theorem 1}$\;\;\;$ {\it For any $k\in \nn$
and $s=1,2,\ldots,k\; ,$ there exists an algorithm $A_s$ for solving 
(\ref{1}) with the right-hand side $f\in F^{r,\rho}$ that computes 
for each $n\in \nn$ and  $\delta\in (0,\frac{1}{2})$ 
an approximation~$l^s$ to the solution $z$, and has the following 
properties.
\f
(i)$\;\;$ There exist  constants $\bar{C}_1^s$ and $C_1^s$, depending only 
on the parameters of the class $F^{r,\rho}$, $a$~and~$b$ (and independent of $\eta$), 
such that    
\be
\sup_{t\in [a,b]} \| z(t)-l^s(t)\|
\leq 
\left\{ \begin{array}{ll} 
\bar{C}_1^s n^{-\alpha_s}\;\;&{\rm \mbox{ if }} L(b-a)
                              {\rm \mbox{ is arbitrary and }} 
			      \; n\geq L(b-a)/\ln 2 \; , \\
C_1^s (b-a)^{q+1}n^{-\alpha_s} \;\;&{\rm \mbox{ if }} L(b-a)\leq \ln 2 
                               {\rm \mbox{ and }} n\in \nn \, ,
\end{array}
\right.
\label{th11}
\ee
where
\be
\alpha_s = \left\{ \begin{array}{ll} q(2^s-1) + 2^{s-1}-1\;\; 
                                            & {\rm \mbox{ {\rm in RAND}}} \\
           qs+s-1 \;\; & {\rm \mbox{ {\rm in QUANT} ,}} \end{array}
	\right. 
	\label{th12}
\ee
and $\;\; \sup \{ C_1^s: \;\; b-a\leq M \} < \infty$ for any $M>0$.
\f
This holds (for $n\geq 5$) with probability at least $1-\delta$.
\vsn
(ii)$\;\;$ There exists a constant $C_2^s$, depending only on the parameters
of the class $F^{r,\rho}$, $a$ and~$b$ (and independent of $\eta$), 
such that for $n\in \nn $ the cost of $A_s$ is bounded by 
\be
{\rm cost}(A_s) 
\leq C_2^s n^{\beta_s}\left( \beta_k \log n +\log \frac{1}{\delta}\right),
\label{th13}
\ee
where
\be
\beta_s = \left\{ \begin{array}{ll} 2^s-1 \;\; 
                                            & {\rm \mbox{ {\rm in RAND}}} \\
           s \;\; & {\rm \mbox{ {\rm in QUANT .}}} \end{array}
	\right. 
	\label{th14}
\ee 
and $\;\; \sup \{ C_2^s: \;\; b-a\leq M \} < \infty$ for any $M>0$.
} 
\vsn
{\bf Proof}$\;\;$ See Section 5. \qed
\vsn
Let us comment on the conditions in Theorem~1 satisfied by the constants 
$C_1^s$ and $C_2^s$. The algorithm $A_s$
will be defined by recursive application of $A_{s-1},\ldots, A_1$ 
on intervals of length going to $0$. The properties of $C_1^s$ and $C_2^s$ 
assure that constants appearing in the error and cost estimates do not grow to infinity 
as the length of the interval tends to $0$.
\vsn
The error and cost estimates for algorithm $A_k$ 
in the randomized and quantum versions, with a~suitably chosen 
parameter $k$, play the main role in establishing the complexity
of problem (\ref{1}).  The crucial point is the 
convergence from above, as $k\to \infty$, of the sequence 
$\{ \beta_k/\alpha_k \}$ to the optimal exponent,
\begin{displaymath}
\lim\limits_{k\to \infty} \frac{\beta_k}{\alpha_k} =
\left\{ \begin{array}{ll} {\displaystyle \frac{1}{q+\frac{1}{2}} } \;\;  & {\rm \mbox{ in RAND}} \\
         {\displaystyle  \frac{1}{q+1} } \;\; & {\rm \mbox{ in QUANT .}} \end{array}
	\right. 
\end{displaymath}
Note that the results for $k=2$ agree with those from \cite{new}, 
since $\beta_2/\alpha_2$ equals to $1/(q+\frac{1}{3})$ in RAND, and 
$1/(q+\frac{1}{2})$ in QUANT.
\f
The following theorem gives essentially optimal 
upper complexity bounds for problem (\ref{1}).
\vsn
{\bf Theorem 2}$\;\;$ {\it For any $\gamma \in (0,1)$, there exist 
positive constants
$C_1(\gamma)$, $C_2(\gamma)$ and $\e_0(\gamma)$ (depending only on $\gamma$,
the parameters of the class $F^{r,\rho}$, $a$ and $b$)
such that for all $\e \in (0, \e_0(\gamma))$  and $\delta\in (0,\frac{1}{2})$
the $\e$-complexity in the randomized and quantum settings satisfies
\be
\comp^{{\rm rand}}(F^{r,\rho},\e) \leq C_1(\gamma) 
\left(\frac{1}{\e} \right)^{1/(q+1/2-\gamma) }
\label{comprand}
\ee
and
\be
\comp^{{\rm quant}}(F^{r,\rho},\e,\delta) \leq 
 C_2(\gamma) 
\left(\frac{1}{\e}\right)^{1/(q+1-\gamma)}\log\frac{1}{\delta}  \, .
\label{compquant}
\ee
}
{\bf Proof}$\;\;$ Define 
\be
k=\left\{ \begin{array}{ll} \lceil \log \left( 1/\gamma+1 \right) 
               \rceil & \mbox{ \rm {in RAND}} \\
               \lceil 2/\gamma \rceil & \mbox{ \rm {in QUANT .}}
	\end{array} \right. 
\label{defk}
\ee	
Consider the randomized setting. We pass from the probabilistic error to the error given by
(\ref{3a}) in a usual way, by selecting a suitable $\delta$. 
Let $K$ denote
a positive (deterministic) upper bound on $e^{\omega}(A_k, f)$,
depending only on the parameters of the class $F^{r,\rho}$, $a$ and $b$.
(One can see from the proof of Theorem 1 that such a bound exists.)
 Then the randomized error (\ref{3a}) of $A_k$ is bounded by
$$
e^{{\rm rand}}(A_k, F^{r,\rho})^2 \leq K^2\delta+ (\bar{C}_1^k n^{-\alpha_k})^2
$$
(for $n$ sufficiently large). Selecting $\delta=3\e^2/(4K^2)$ and 
$n=\lceil (2\bar{C}_1^k/\e)^{1/\alpha_k} \rceil$, we get that
$e^{{\rm rand}}(A_k, F^{r,\rho}) \leq \e$. The cost of $A_k$ is then
bounded by
$${\rm cost}(A_k) = O \left( \left(\frac{1}{\e} \right)^{\beta_k/\alpha_k} 
                      \log\frac{1}{\e} \right), $$
with the constant in 
the $O$-notation depending only  on $k$ and the parameters of the class 
$F^{r,\rho}$, $a$ and $b$. 
By the definition of $k$, we have that
$$\frac{\alpha_k}{\beta_k}\geq q+\frac{1}{2} -\frac{\gamma}{2}.$$ 
Hence, 
$${\rm cost}(A_k) \leq \hat{C}_1(\gamma) 
\left(\frac{1}{\e} \right)^{1/(q+1/2-\gamma/2)} \log \frac{1}{\e}
\leq C_1(\gamma) 
\left(\frac{1}{\e}\right)^{1/(q+1/2-\gamma)}, $$ 
for suitable constants $\hat{C}_1(\gamma)$, $C_1(\gamma) $, and
$\e$ sufficiently small. This proves (\ref{comprand}). 
\f
We handle the quantum setting similarly.  
We estimate the error $e^{{\rm quant}}(A_k, F^{r,\rho},\delta)$
using (\ref{th11}) and (\ref{th12}), select $n$ to assure that
$e^{{\rm quant}}(A_k, F^{r,\rho},\delta)\leq \e$, and use the cost bound
(\ref{th13}). The choice (\ref{defk}) of $k$ leads to (\ref{compquant}).\qed
\vsn
Comparing upper bounds from Theorem 2 with lower bounds given 
in (\ref{rec2}) and (\ref{rec4}),
we see that they match up to a small parameter $\gamma$ in the exponent. In the quantum setting
this holds up to a logarithmic factor that depends on $\delta$.  
\vsn
{\Large \section{ Algorithms in the randomized and quantum settings }}
\noindent
We define algorithms whose properties 
are the subject of Theorem 1. The definition is recursive.
In both the randomized and quantum settings, we inductively define 
a sequence of algorithms $A_1,A_2,\ldots, A_s,\ldots, A_k$
for solving problem (\ref{1}). The algorithm $A_k$, with a properly 
chosen index $k$, will be our final algorithm.
\vsn
The points $\{x_i\}$ define the uniform partition of $[a,b]$, 
$\; x_i=a+ih$, $i=0,1,\ldots,n$ with $h=(b-a)/n$. We shall call $n$ 
a {\it basic parameter}. To display the dependence on the interval
over which the algorithm is applied and on the basic parameter used, we shall use the notation 
$A_s=A_s([a,b],n)$.  An approximation to the solution $z$ computed in
$A_s([a,b],n)$ is denoted by $l^s=l^s(t), \;\; t\in [a,b].$
\vsn
We start the induction with $s=1$. The algorithm $A_1([a,b],n)$ 
is defined to be Taylor's
algorithm in $[a,b]$ with step size $h$. We set $y_0=\eta$.
For a given $y_i$ (an approximation to $z(x_i)$), 
we let $z_{i} $ be the solution of the local problem
\be
z_{i} '(t)= f(z_{i}(t)), \;\; t\in [x_i, x_{i+1}] \; , \;\;\;\;\;
z_{i}(x_i) = y_i \, ,
\label{p1}
\ee
and we define $y_{i+1} = l^0_{i}(x_{i+1})$. Here, 
$l_{i}^0(t)$ is the truncated Taylor's expansion of $z_i$ 
for $t\in [x_i, x_{i+1}] $, given by
$$l_{i}^0(t)=\sum\limits_{j=0}^{r+1} (1/j!) 
z^{(j)}_{i}(x_i)(t-x_i )^j.$$ 
The approximation to $z$ in  $A_1([a,b],n)$ is defined by 
\be
l^1(t)= l_i^0(t) \;\;\; \mbox{ for } t\in [x_i, x_{i+1}].
\label{p3}
\ee
Suppose that the algorithm $A_s([a,b],n)$ is defined for any 
$[a,b]$ and $n$. 
We now describe how to get $A_{s+1}([a,b],n)$ from $A_{s}([a,b],n)$. 
\f
We first inductively define a sequence of approximations 
$\{y_i\}$ in $A_{s+1}([a,b],n)$, where 
$y_i \approx z(x_i)$, $i=0,1,\ldots,n$. We set $y_0=\eta$. For a given $y_i$, 
we apply the algorithm $A_s([x_i,x_{i+1}],m)$ to the problem (\ref{p1}), with
a certain basic parameter $m$ to be chosen later on. The approximating
function obtained in this way is denoted by 
$l_i^s=l_i^s(t)$, $t\in [x_i,x_{i+1}]$.
\f
The crucial point, which we now describe, is how we obtain $y_{i+1}$ from $y_i$ 
and $l_i^s$. We divide the interval $[x_i,x_{i+1}]$ into
$ml$ subintervals of equal length,  with end points
$$ z_j^i= x_i + j \bar{h}, $$
with $ j=0,1,\ldots, ml, \,$ where $\, \bar{h}=(x_{i+1}-x_i)/(ml)$. 
Here $l$ is another parameter that will be chosen in the sequel. 
(The notation $l$ used for the parameter will not be confused with 
$l^s=l^s(t)$ denoting the approximating function, since the latter always appears 
with a superscript. The~distinction is clear from the context.)
\f
The solution of (\ref{p1}) satisfies the identity
$$
z_i(x_{i+1})=z_i(x_i) + \sum\limits_{j=0}^{ml-1} 
\int\limits_{z_j^i}^{z_{j+1}^i} f(z_i(t))\, dt \, .
$$
For $j=0,1,\ldots,ml-1$ and $i=0,1,\ldots, n-1$, define the polynomial
\be
w_{ij}(y) = \sum\limits_{k=0}^r \frac{1}{k!} f^{(k)}(l_i^s(z_j^i))
(y-l_i^s(z_j^i))^k \, .
\label{p5}
\ee
Then we can equivalently write
\be
\begin{array}{lll}
z_i(x_{i+1})&=&{\displaystyle z_i(x_i)+ \sum\limits_{j=0}^{ml-1} 
                     \int\limits_{z_j^i}^{z_{j+1}^i} w_{ij}(l_i^s(t))\, dt } \\
         &+& {\displaystyle \bar{h}^{q+1} ml\, \frac{1}{ml} 
        	 \sum\limits_{j=0}^{ml-1} \int\limits_{0}^{1} 
               g_{ij}(u) \, du }\\
        &+& {\displaystyle \sum\limits_{j=0}^{ml-1} \int\limits_{z_j^i}^{z_{j+1}^i} 
         \left( f(z_i(t))- f(l_i^s(t)) \right)\, dt} \, ,
\end{array}
\label{ident1}
\ee
where function $g_{ij}$ is defined for $u\in [0,1]$ by 
\be
g_{ij}(u)= \frac{  f(l_i^s(z_j^i +u\bar{h}))- w_{ij}(l_i^s(z_j^i +u\bar{h})) }
{ \bar{h}^q}\, .
\label{gij}
\ee
(The value $l_i^s(z_{j+1}^i)$ is meant here as the limit $\lim \, l_i^s(t)$
as $t\to z_{j+1}^i$ from the left.)
\f
To get the formula for $y_{i+1}$, we neglect 
the last right-hand side term in (\ref{ident1}), 
and we approximate the penultimate term. We arrive at the formula
\be
\begin{array}{lll}
y_{i+1}&=&{\displaystyle y_i + \sum\limits_{j=0}^{ml-1} \int\limits_{z_j^i}^{z_{j+1}^i} 
w_{ij}(l_i^s(t))\, dt} \\
         &+&{\displaystyle  \bar{h}^{q+1} ml\, {\rm AP}_i(f)} \, .
\end{array}
\label{yi}
\ee
The vector ${\rm AP}_i(f)$  is constructed to  approximate the mean 
of $ml$ integrals
\be
{\rm AP}_i(f) \approx \frac{1}{ml}  \sum\limits_{j=0}^{ml-1} \int\limits_{0}^{1} 
g_{ij}(u) \, du 
\label{Ai}
\ee
as follows. 
\f
Approximate the integral of $g_{ij}$ by the mid-point rule with $N$ knots $u_k$, i.e.,
\be
\int\limits_0^1 g_{ij}(u)\, du \approx 
\frac{1}{N} \sum\limits_{k=0}^{N-1} g_{ij}(u_k)\, .
\label{p11}
\ee
Hence,
\be
\frac{1}{ml} \sum\limits_{j=0}^{ml-1} 
\int\limits_0^1 g_{ij}(u)\, du \approx 
\frac{1}{mlN} \sum\limits_{j=0}^{ml-1} \sum\limits_{k=0}^{N-1} g_{ij}(u_k).
\label{c11}
\ee
The (random) vector ${\rm AP}_i(f)$ is computed, in the appropriate setting,
by applying the optimal randomized or quantum algorithm (with 
repetitions) to approximate each component of 
the mean of $mlN$ vectors in the right-hand side of (\ref{c11}). 
For a discussion of algorithms for computing the mean, 
see \cite{Heinrich}, \cite{HN}.
The approximation error is required to be 
bounded by $\e_1$,
\be
\bigg{\|} {\rm AP}_i(f)-\frac{1}{mlN} \sum\limits_{j=0}^{ml-1} 
   \sum\limits_{k=0}^{N-1} g_{ij}(u_k) \bigg{\|} \leq \e_1,
\label{d11}
\ee
with probability at least $1-\delta_1$.
The parameters $N$, $\e_1$ and $\delta_1$ are to be chosen, and 
will be defined soon. 
\f
An approximation $l^{s+1}$ to $z$ over $[a,b]$ in the algorithm 
$A_{s+1}([a,b],n)$ is defined by $l^{s+1}(t)=l_i^s(t)$ for 
$t\in [x_i,x_{i+1})$, and $l^{s+1}(b)=l_{n-1}^s(b)$. 
This completes the definition of $A_{s+1}$ and, by induction, 
of the sequence of algorithms $\{A_s([a,b],n)\}_{s\geq 1}$. 
\vsn
It remains to define the parameters $m$, $l$, $N$, $\e_1$ and $\delta_1$.
They are selected to assure  the best performance of the 
algorithms, and  result from
solving a number of auxiliary optimization problems. 
We do not present here the rather technical analysis 
leading to the proper selection. We give instead  
the resulting values of the parameters, and prove  in the next sections 
the correctness of our choice. To accomplish the definition of algorithm $A_{s+1}$
we make the following choice of parameters: 
\be
(m,l,N) = \left\{ \begin{array}{ll} 
                  (n^2, n^{2^{s+1}-4}, n^{2^s-1})& \;\;\;\;\; 
		  {\rm in} \; {\rm RAND}\\
		  (n,n^{s-1},n^s) & \;\;\;\;\; {\rm in} \;{\rm QUANT} \, , 
\end{array} \right.
\label{parameters}
\ee
and we set $\e_1=1/N$. The parameter $\delta_1$ is chosen as a function of 
$\delta$, the basic parameter $n$ and $k$ to be
\be
\delta_1 = \left\{ \begin{array}{ll} 
                  1-(1-\delta)^{1/n^{2^k-1}}  & \;\;\;\;\; 
		  {\rm in} \; {\rm RAND}\\
		  1-(1-\delta)^{1/n^k} & \;\;\;\;\; {\rm in} \;
		                    {\rm QUANT} \, .
\end{array} \right.
\label{delta11}
\ee
It is useful to illustrate with an example the choice of the basic parameter
in the recursive procedure described above. Consider the randomized case with $s=3$.
We have that $m=n^2$, i.e., at each step 
a new basic parameter is obtained by squaring the current one.
The algorithm $A_3([a,b],n)$ is thus defined by $n$ applications of $A_2([x_i,x_{i+1}],n^2)$
for $0\leq i\leq n-1$. Computing   $A_2([x_i,x_{i+1}],n^2)$
requires in turn applications of $A_1$ on subintervals 
of length $(x_{i+1}-x_i)/n^2 = (b-a)/n^3$, with the basic parameter $m=(n^2)^2 = n^4$.
At the final step, the function $l^3$ is given by Taylor's approximations (polynomials)
on subintervals of length $(b-a)/(n^3m) =  (b-a)/n^7$. (For an arbitrary $s$, intervals in which
$l^s$ is a polynomial will be established in the next section.)
\f
Let us remark that the algorithms $A_1$ and $A_2$ defined above are the well known Taylor algorithm
and the algorithm analyzed in \cite{new}, respectively. For $k\geq 3$, algorithms 
$A_k$ are new. 
\f
{\Large \section{ Performance analysis  }}
\noindent
This section is devoted to the proof of Theorem 1. We first show 
error bounds for the algorithm~$A_s$, neglecting all probabilistic 
considerations related to the bounds. Such issues are  considered in the next subsection, 
as well as the cost of the algorithm.
\f
\subsection{Error bounds}
\noindent
We shall prove (\ref{th11}). To derive error bounds, 
we first discuss the regularity of the 
approximation $l^s$ computed in $A_s([a,b],n)$, and the regularity of the function
$g_{ij}$ given by (\ref{gij}). Note that the 
function $l^s$ is in general not continuous. It is a polynomial on
each subinterval $[c,p)$ of a uniform partition of $[a,b]$ (the last subinterval ending with
$b$ is a closed one),  with length 
\be
(b-a)/n^{2^s-1} \;\; \mbox{{\rm in RAND}}
                   \;\; {\rm and} \;\;
		   (b-a)/n^s  \;\; \mbox{{\rm in QUANT}}.
\label{dlugosc}
\ee
To prove this, note that the length of each  subinterval in which
the approximation $l^1$ in $A_1([a,b],n)$ is a polynomial 
equals to $(b-a)/n$. Let the length 
of each subinterval in which $l^s$ in $A_s([a,b],n)$ is a polynomial 
be equal to $(b-a)/n^{\phi_s}$. From the definition, the function $l_i^s$ 
is a polynomial on each subinterval of length 
\be
(b-a)/(nm^{\phi_s}) = \left\{
\begin{array}{ll} (b-a)/n^{1+2\phi_s} & \;\; \mbox{{\rm in RAND}}\\
		   (b-a)/n^{1+\phi_s} & \;\; \mbox{{\rm in QUANT}} .
\end{array}
\right.
\label{dlugosc1}
\ee
Since functions $l_i^s$ define $l^{s+1}$, we get the recurrence 
relation 
\be
\phi_{s+1}=  \left\{
\begin{array}{ll} 1+2\phi_s  & \;\; \mbox{{\rm in RAND}}\\
		   1+\phi_s & \;\; \mbox{{\rm in QUANT}} 
\end{array}
\right.
\label{dlugosc2}
\ee
with $\phi_1 =1$. This gives the desired statement (\ref{dlugosc}).
\f
Function $l_i^s$ is thus a polynomial 
on each interval $[z_j^i, z_{j+1}^i)$, see (\ref{ident1}), (\ref{gij})
and (\ref{yi}). 
We use this property to observe that the following fact concerning $g_{ij}$ holds.
\vsn
{\bf Fact 1}$\;\;$ {\it There exist positive constants $M_1$ and $M_2$,
depending only on the parameters of the class $F^{r,\rho}$, $a$ and $b$ 
(and independent of $i$, $j$, $z_j^i$, $y_j^i$), such that
\be 
\|g_{ij}(u)\| \leq M_1, \;\;\; u\in [0,1],
\label{g1}
\ee
\be
\|g_{ij}(u_1) - g_{ij}(u_2)\| \leq M_2 
|u_1 - u_2|, \;\;\; u_1, u_2\in [0,1]\, ,
\label{g2}
\ee
and $\; \sup\{\; \max\{M_1,M_2\}:\;\;\; b-a\leq M\; \}\, <\, \infty$ for any 
$M>0$.
   }
\vsn  
{\bf Proof}$\;\;$ The proof follows from arguments 
used in the proof of Lemma on p.~828 in \cite{rand5}. We replace 
$h:=\bar{h}$, $x_i:=z_j^i$, $x_{i+1}:=z_{j+1}^i$, $l_i^*:=l_i^s$,  
$y_i^*:=l_i^s(z_j^i)$ and $w_i^*:= w_{ij}$ in that Lemma. We use the fact that function $l_i^s$ 
is a polynomial in $[z_j^i, z_{j+1}^i)$ of the form 
$$
l_i^s(t)= \sum\limits_{k=0}^{r+1} (1/k!) z_{ij}^{(k)} (z_j^i) (t-z_j^i)^k,
$$
where $z_{ij}$ satisfies the equation $z_{ij}'(t)= f(z_{ij}(t))$, $t\in [z_j^i,z_{j+1}^i)$, with
the initial condition $z_{ij}(z_j^i)= c$ resulting from the recursive definition of the 
algorithms. \qed
\vsn
We are ready to prove 
(\ref{th11}). We proceed by induction. Let $s=1$. The desired bound 
in this case is a version of 
the well known error estimate for Taylor's method. The error of Taylor's 
expansion formula satisfies 
$$
\|z_i(t)-l_i^0(t)\| \leq \hat{M} (t-x_i)^{q+1},
$$
for $t\in [x_i,x_{i+1}]$, where $\hat{M}$ depends only on the parameters 
of the class
$F^{r,\rho}$, $a$, $b$ and is bounded for bounded $b-a$. 
First estimating the error $\|z(x_i)-y_i\|$ in the usual way,
and then passing to the error over $[a,b]$, we arrive at the bounds
$$
\sup_{t\in [a,b]} \| z(t)-l^1(t)\|
\leq 
\left\{ \begin{array}{ll} 
e^{Lh}(e^{L(b-a)} -1)L^{-1} \hat{M} h^q + \hat{M} h^{q+1}  \;\;&
{\rm \mbox{ if }} L(b-a) {\rm \mbox{ is arbitrary, }} \\
5\hat{M} (b-a)h^{q} \;\;&{\rm \mbox{ if }} L(b-a)\leq \ln 2 \, .
\end{array}
\right.
$$
Hence, (\ref{th11}) holds with 
$$
\bar{C}_1^1=2 (e^{L(b-a)} -1)L^{-1} \hat{M} (b-a)^q + 
\hat{M}(b-a)^{q+1},\;\;\;\;
C_1^1=5\hat{M}, \;\;\;\; {\rm \mbox{and}}\;\;\;\; \alpha_1=q. 
$$
Suppose by induction that (\ref{th11}) holds for $l^s$, 
for any $[a,b]$ and $n$. 
Let $e_i=z(x_i)-y_i$. By the triangle inequality, 
\be
\|e_{i+1}\| \leq \|z(x_{i+1}) - z_i(x_{i+1})\| + \|z_i(x_{i+1}) - y_{i+1}\|.
\label{bl1}
\ee
From the dependence 
of the solution on initial conditions, we see that the first term is bounded by 
\be
\|z(x_{i+1}) - z_i(x_{i+1})\|\leq e^{Lh} \|e_i\|.
\label{bl2}
\ee
To estimate the second term, we subtract (\ref{yi}) from (\ref{ident1}) to get
\be
\begin{array}{lll}
z_i(x_{i+1}) - y_{i+1}& = &{\displaystyle \bar{h}^{q+1} ml \left( \frac{1}{ml} 
        	 \sum\limits_{j=0}^{ml-1} \int\limits_{0}^{1} 
               g_{ij}(u) \, du -  {\rm AP}_i(f) \right) } \\
       & + &{\displaystyle  \sum\limits_{j=0}^{ml-1} \int\limits_{z_j^i}^{z_{j+1}^i} 
         \left( f(z_i(t))- f(l_i^s(t)) \right)\, dt \, }.
\end{array}
\label{bl3}
\ee
From the definition of ${\rm AP}_i(f)$, and the properties of $g_{ij}$ 
stated in Fact 1, we write the bound
\be
\bigg{\|} \frac{1}{ml} \sum\limits_{j=0}^{ml-1} \int\limits_{0}^{1} 
               g_{ij}(u) \, du -  {\rm AP}_i(f) \bigg{\|} \leq CN^{-1} + \e_1 .
\label{bl4}
\ee
Here, the term $CN^{-1}$ is the estimate of the mid-point rule error, 
and $\e_1$ comes from the randomized or quantum approximation 
of the mean of $mlN$ vectors, see (\ref{d11}).  
(The last bound holds with probability at least $1-\delta_1$.)
The constant $C$ depends only on the parameters of the class $F^{r,\rho}$, 
$a$ and $b$, and is bounded for bounded $b-a$. 
\f
The second right-hand side term in (\ref{bl3}) will be estimated from
the definition of $l_i^s(t)$. 
\f
Let $n\geq L(b-a)/\ln 2$. Then 
$L(x_{i+1}-x_i) \leq \ln 2$, so that the second inequality 
in (\ref{th11}) holds for $A_s([x_i,x_{i+1}],m)$ (with $m=n^2$ in RAND and $m=n$
in QUANT).
We have 
\be
\| f(z_i(t))- f(l_i^s(t)) \| \leq L\| z_i(t) - l_i^s(t)\| \leq 
LC_1^s(x_{i+1}-x_i)^{q+1} m^{-\alpha_s}.
\label{bl5}
\ee
(The coefficient $C_1^s$ for $A_s([x_i,x_{i+1}],m)$ depends on 
$x_i$, $x_{i+1}$, but, due to the inductive assumption on $C_1^s$, 
it is bounded by a constant depending only on the parameters of the class 
$F^{r,\rho}$, $a$ and $b$. We denote the bound by the same symbol
$C_1^s$.)
\f
Putting together relations (\ref{bl2})--(\ref{bl5}), we can use (\ref{bl1}) to see that 
\be
\begin{array}{ll}
\|e_{i+1}\| &{\displaystyle \leq e^{Lh} \|e_i\| + \bar{h}^{q+1} ml \left(CN^{-1}+\e_1\right)}\\
            &{\displaystyle + LC_1^s(b-a)^{q+2} n^{-(q+2)} m^{-\alpha_s} }\, ,
\end{array}
\label{bl6}
\ee
for $i=0,1,\ldots, n-1$, where $e_0=0$.
Hence, 
\be
\|e_i\| \leq \frac{e^{Lhn} -1}{e^{Lh} -1}\left(
\bar{h}^{q+1} ml \left(CN^{-1}+\e_1\right)
            + LC_1^s(b-a)^{q+2} n^{-(q+2)} m^{-\alpha_s}\right)\, ,
\label{bl7}
\ee 
for $i=0,1,\ldots,n$. 
\f
We now pass to the error in $[a,b]$. For $t\in [x_i,x_{i+1}]$, we have 
\be
\begin{array}{ll}
\|z(t)-l^{s+1}(t)\|&=\|z(t)-l_i^s(t)\| \leq \|z(t)-z_i(t)\| + 
                                     \|z_i(t)-l_i^s(t)\| \\ 
                    &\leq e^{Lh} \|e_i\| + 
    C_1^s h^{q+1} m^{-\alpha_s} \, .
 \end{array}
\label{bl8}
\ee
We now use (\ref{bl7}) in (\ref{bl8}). Since $e^{Lh}-1\geq Lh$, 
we get for arbitrary $L(b-a)$ and $n\geq L(b-a)/\ln 2$ that 
\be
\begin{array}{ll}
\sup\limits_{t\in [a,b]} \|z(t)-l^{s+1}(t)\| &\leq 2\left( e^{L(b-a)}-1 \right)
                                        L^{-1}(b-a)^q
                                      (mln)^{-q}(CN^{-1}+\e_1) \\
&+\left( 2 e^{L(b-a)}-1\right) C_1^s (b-a)^{q+1} n^{-(q+1)}
m^{-\alpha_s} \, .
\end{array}
\label{bl9}
\ee
If the interval $[a,b]$ is small enough that $L(b-a)\leq \ln 2$, then
${\rm exp}(L(b-a)) -1\leq 2L(b-a)$, and we obtain
\be
\begin{array}{ll}
\sup\limits_{t\in [a,b]} \|z(t)-l^{s+1}(t)\| &\leq 4 (b-a)^{q+1}
                                      (mln)^{-q}(CN^{-1}+\e_1) \\
&+\left( 4L(b-a) +1\right) C_1^s (b-a)^{q+1} n^{-(q+1)} m^{-\alpha_s} \, .
\end{array}
\label{bl10}
\ee
Bounds (\ref{bl9}) and (\ref{bl10}) allow us to finish the inductive proof 
of (\ref{th11}). To close the induction, we continue with two cases, randomized and quantum,
introducing proper parameters $m$, $l$, $N$ and~$\e_1$ given in 
(\ref{parameters}). 
\f
Consider the randomized setting. Introducing the parameters we get
\be
\sup_{t\in [a,b]} \| z(t)-l^{s+1}(t)\| \leq 
\left\{ \begin{array}{ll} 
\bar{C}_1^{s+1} n^{-\alpha_{s+1}}\;\;&{\rm \mbox{ if }} L(b-a)
                              {\rm \mbox{ is arbitrary and }} 
			      \; n\geq L(b-a)/\ln 2 \; , \\
C_1^{s+1} (b-a)^{q+1}n^{-\alpha_{s+1}} \;\;&{\rm \mbox{ if }} L(b-a)\leq \ln 2 
                               {\rm \mbox{ and }} n\in \nn \, ,
\end{array}
\right.
\label{th111}
\ee
where 
\be
\alpha_{s+1}= \min\{\, (2^{s+1}-1)q+2^s-1\, ,\, q+1+2\alpha_s \, \}\, .
\label{th121}
\ee
Constants $\bar{C}_1^{s+1}$ and  $C_1^{s+1}$ are given by
\be
\bar{C}_1^{s+1} = (b-a)^q \left( 2\left( e^{L(b-a)}-1 \right) L^{-1} (C+1)
+ \left(2e^{L(b-a)}-1\right) C_1^s (b-a) \right)
\label{barC}
\ee
and
\be
C_1^{s+1} = 4(C_1^s+C+1)\, .
\label{C}
\ee
Since $\alpha_s$ is given by (\ref{th12}), the same holds 
due to (\ref{th121}) for $\alpha_{s+1}$,
and the induction is closed in the randomized setting.
\f
In the quantum setting, after introducing the parameters from 
(\ref{parameters}), we arrive at the bound (\ref{th111}), where only
the exponent $\alpha_{s+1}$ differs from that in the randomized setting.
It is now given by
\be
\alpha_{s+1}= \min\{\, (s+1)q+s\, ,\, q+1+\alpha_s \, \}\, .
\label{th123}
\ee
Constants $\bar{C}_1^{s+1}$ and  $C_1^{s+1}$ remain the same as in
the randomized setting, see (\ref{barC}) and  (\ref{C}). 
\f
Since $\alpha_{s+1}$ agrees with (\ref{th12}), 
the inductive proof of (\ref{th11}) is completed in the 
quantum setting.
\f
\subsection{Probability and cost}
\noindent
In the previous subsection we neglected  probability 
issues related
to the bounds obtained. The cost of computations was also not 
considered. We now show that (\ref{th11}) holds with probability
at least $1-\delta$, and we prove the cost bound (\ref{th13}).
\f
We first show that 
\be
{\rm \mbox{bound (\ref{th11}) holds with probability at least}} 
\;(1-\delta_1)^{\psi(n,s)},
\label{d1}
\ee
where $\psi (n,1) =0$ for all $n$, and 
\be
\psi (n,s)=\left\{ \begin{array}{ll} {\displaystyle \sum\limits_{i=1}^{s-1} n^{2^i-1}  } \;\; 
                                            & {\rm \mbox{ in RAND}} \\
          {\displaystyle  \frac{n^s-n}{n-1} } \;\; & {\rm \mbox{ in QUANT }} 
                    \end{array} 
            \right. 
\label{pr1}
\ee
for $s\geq 2$ and $n\geq 2$.
Indeed, for $s=1$ the error bound for Taylor's algorithm holds with certainty, 
that is, $\psi (n,1) =0$~. For $s=2$, (\ref{th11}) holds with probability at least
$(1-\delta_1)^n$, so that $\psi(n,2)=n$, which agrees with (\ref{pr1}). 
Suppose that (\ref{d1}) and (\ref{pr1}) hold for some $s\geq 2$.
The error estimate for $l^{s+1}$ holds true if the error bounds for $l_i^s$ 
and bounds (\ref{d11}) are satisfied 
for $i=0,1,\ldots,n-1$. This holds with probability at least
$(1-\delta_1)^{n\psi(m,s)+n}$. Hence, we have that
\be
\psi(n,s+1)=n\psi(m,s)+n.
\label{pr2}
\ee
This yields that $\psi(n,s+1)$ satisfies (\ref{pr1}), which follows after introducing
$m=n^2$ in the randomized case, and $m=n$ in the quantum case. 
\f
Note that for $n\geq 5$, we have 
$$
\psi(n,s)\leq \left\{ \begin{array}{ll}  n^{2^s-1}\;\; 
                                            & {\rm \mbox{ in RAND}} \\
           n^s \;\; & {\rm \mbox{ in QUANT .}} 
                    \end{array} 
            \right. 
$$
This, and the choice of $\delta_1$ given in (\ref{delta11}), assures that 
$$
(1-\delta_1)^{\psi(n,s)} \geq 1-\delta 
$$
for $s=1,2,\ldots,k$, as claimed.
\f
We now prove the cost estimate (\ref{th13}) by induction. For $s=1$, 
the cost is bounded by $c_1(r,d)n$ classical evaluations of components of 
$f$ or its partial derivatives, where $c_1(r,d)$ depends only 
on $r$ and $d$. Hence, (\ref{th13}) holds with $\beta_1=1$.
\f
Suppose that (\ref{th13}) is satisfied for $A_s$. The costs of computing
successive components that define the approximation $l^{s+1}$ in $A_{s+1}$ 
are bounded as follows.
The cost of computing $l_i^s$ for all $i$ is bounded by 
$$C_2^s nm^{\beta_s} \left(\beta_k\log m + \log \frac{1}{\delta} \right),$$
and the cost of $w_{ij}$ for all $i$, $j$ by 
$$c_2(r,d)nml,$$ 
where
$c_2(r,d)$ depends only on $r$ and $d$. Here, $C_2^s$ can be chosen 
to depend only on the parameters of the class  $F^{r,\rho}$, 
$a$ and $b$, and is bounded for bounded $b-a$. 
\f
It remains to take into account the cost of computing 
${\rm AP}_i(f)$, $i=0,1,\ldots,n-1$. We shall use the results
on computation of the mean of real numbers,
those of Math\'e~\cite{HN} in the randomized setting, and of Brassard et al.~\cite{Bras} 
in the quantum setting. 
Let $\kappa=2$ in RAND, and $\kappa=1$ in QUANT. 
To compute  ${\rm AP}_i(f)$ for $i=0,1,\ldots,n-1$, each satisfying (\ref{d11}), 
we need at most
$$
Pn \min \left\{  \, mlN, \left(\frac{1}{\e_1} \right)^{\kappa}\, \right\} 
\log \frac{1}{\delta_1}
$$
evaluations of $f$ in RAND, or quantum queries on $f$ in QUANT. 
The constant $P$ 
depends only on the parameters of the class $F^{r,\rho}$, $a$ and $b$,
and is bounded for bounded $b-a$. The logarithmic factor 
is related to the number of repetitions needed to increase 
probability of (\ref{d11}). For the basic algorithm for computing 
the mean, 
(\ref{d11}) holds with probability at least $\frac{3}{4}$, and for our
purposes it must be at least $1-\delta_1$. 
For a discussion of the role of repetitions and computing the median
of a number of results,  see  \cite{Heinrich}. 
\f
Putting together these bounds, we get
\be
{\rm cost}(A_{s+1}) 
\leq C_2^s n m^{\beta_s}\left( \beta_k \log m +\log \frac{1}{\delta} \right)
+c_2(r,d)nml+ Pn \min \left\{ \, mlN, \left(\frac{1}{\e_1}\right)^{\kappa}\, \right\}
\log \frac{1}{\delta_1}  .
\label{co1}
\ee
Note that
$$
\log \frac{1}{\delta_1} = \log \frac{1}{1-(1-\delta)^{1/n^{\beta_k}}}  \leq
c \left(\log \frac{1}{\delta} + \beta_k \log n \right),
$$
where $c$ is an absolute constant. 
\f
Finally, we introduce in (\ref{co1}) the parameters $m$, $l$, $N$ 
and $\e_1$ defined by (\ref{parameters}). 
\f
In the randomized case,  we arrive at 
\be
{\rm cost}(A_{s+1})  \leq 
C_2^{s+1} n^{\beta_{s+1}}\left( \beta_k \log n +\log \frac{1}{\delta} \right),
\label{co2}
\ee
where $\beta_{s+1}=\max \{\, 2\beta_s +1, 2^{s+1}-1\, \}$ (which 
agrees with (\ref{th14})), and $C_2^{s+1}= 2C_2^s +c_2(r,d) +Pc$. 
\f
In the quantum case, (\ref{co2}) holds with 
$\beta_{s+1}=\max \{\, \beta_s +1, s+1\, \}$ (which again
agrees with (\ref{th14})) and $C_2^{s+1}= C_2^s +c_2(r,d) +Pc$. 
\f
The inductive proof of (\ref{th13}) is thus completed. This also ends the proof
of Theorem 1. 
\vsn
{\Large\section{ Final remarks }}
\noindent
We have established, up to arbitrarily small $\gamma$ in the exponent and a logarithmic factor,
the randomized and quantum complexity of initial-value problems. The matching 
upper bound is obtained by defining
a sequence of randomized and quantum algorithms for solving the problem. The main issues here
are the following:
\begin{itemize}
\item
the use of integral identity (\ref{ident1}) satisfied by the solution, 
\item
proper application of randomized and quantum algorithms for computing the mean of real numbers,
and 
\item
the recursive definition of the algorithms. To compute the approximation at the final level~$k$,
we pass a number of times through all the proceeding levels.
\end{itemize}
\noindent
Let us summarize the results in various settings. 
Neglecting $\gamma$ and the logarithmic factor, we have that the complexity is
of order $(1/\e)^{1/(q+1)}$ in the worst-case deterministic setting with linear information 
\cite{przegl}, and the same holds in the quantum case. In the worst-case deterministic setting with 
standard information, it is of order $(1/\e)^{1/q}$, see~\cite{przegl}. 
In the randomized setting, we stay between these results with complexity
of order $(1/\e)^{1/(q+1/2)}$. 

\end{document}